\documentclass[aps,prb,twocolumn,showpacs,superscriptaddress]{revtex4-1}
\usepackage{amsmath}
\usepackage{epsf}
\usepackage{epsfig}
\usepackage[T1]{fontenc}
\usepackage[latin9]{inputenc}
\usepackage{array}
\usepackage{multirow}

\begin{document}

\title{Carbon chains grown perpendicularly on graphene}

\author{C. Ataca}
\affiliation{Department of Physics, Bilkent University, Ankara
06800, Turkey}
\affiliation{UNAM-Institute of Materials Science and
Nanotechnology, Bilkent University, Ankara 06800, Turkey}
\author{S. Ciraci}\email{ciraci@fen.bilkent.edu.tr}
\affiliation{Department of Physics, Bilkent University, Ankara
06800, Turkey}
\affiliation{UNAM-Institute of Materials Science and
Nanotechnology, Bilkent University, Ankara 06800, Turkey}

\date{\today}

\begin{abstract}

Based on first-principles calculations we predict a peculiar growth process, where carbon adatoms adsorbed to graphene readily diffuse above room temperature and nucleate segments of linear carbon chains attached to graphene. These chains grow longer on graphene through insertion of carbon atoms one at a time from the bottom end and display a self-assembling behavior. Eventually, two allotropes of carbon, namely graphene and cumulene are combined to exhibit important functionalities. The segments of carbon chains on graphene become chemically active sites to bind foreign atoms or large molecules. When bound to the ends of carbon chains, transition metal atoms, Ti, Co and Au, attribute a magnetic ground state to graphene sheets and mediate stable contacts with interconnects. We showed that carbon chains can grow also on single wall carbon nanotubes.

\end{abstract}

\maketitle

\section{Introduction}

\begin{center}
\begin{figure}
\includegraphics[width=8.25cm]{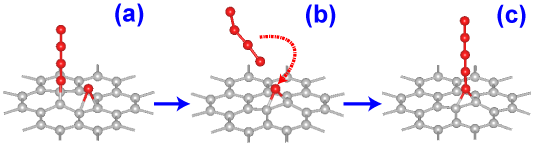}
\caption{(Color online) Schematic description of the growth of a segment of CAC on graphene. (a) A carbon adatom and perpendicularly attached CAC of four carbon atoms (red/dark balls) on graphene surface (honeycomb with grey/light balls). (b) When the carbon adatom becomes within a threshold distance, the bridge bonds of CAC with graphene are broken. (c) Concomitantly, carbon atom at the lower end of CAC is rebound to the adatom through a concerted process. Eventually, CAC becomes longer and has five carbon atoms. }
\label{fig:1}
\end{figure}
\end{center}

Graphene\cite{gra1,gra2}, a strictly two-dimensional allotrope of carbon, has a planar honeycomb structure, that underlies a number of exceptional properties. A segment of carbon atomic chain (CAC), a strictly one-dimensional allotrope, is characterized with its high strength, linear geometry and even-odd disparity occurring in its structural, quantum transport and magnetic properties. CACs have been explored theoretically for a long time\cite{abdul,sefaet,yumur}, and synthesized only recently.\cite{eisler,iijima,naturechem} Here, we portend a unique growth process of CACs on graphene: When two carbon atoms adsorbed on graphene are at close proximity, the potential barrier between them collapses and they form C$_2$ attaching perpendicularly to graphene. A CAC can continue to grow longer on graphene through insertion of carbon atoms one at a time from the bottom end as described in Fig.\ref{fig:1}(a)-(c). This process leads to a number of unusual artificial structures combined of the two allotropes of carbon, namely graphene and CACs. Graphene sheets with protruding CACs can achieve chemical activity and attain new functionalities through CACs capped by foreign atoms or other graphene sheets. A single hydrogen molecule readily dissociates, once it is attached to the top of a CAC. This self-assembling behavior of carbon adatoms can also be exploited for the synthesis of free carbon atomic chains and other artificial nanostructures promising important applications, such as a medium of high capacity hydrogen storage. That the binding energy of a single carbon adatom on graphene is smaller than the cohesive energy of a linear carbon chain underlies the present self-assembling growth process.

The $sp^D$-hybrid orbitals are indigenous to the dimensionality (D=1,2,3) of these allotropic forms. The $sp^2$-bonding together with $\pi$-bonding assures the planar stability of honeycomb structure of graphene. Covalent bonding of $sp^{D=1}$ hybrid orbitals along the chain axis together with $\pi$-bonding of perpendicular $p_x$ and $p_y$ orbitals are responsible for the high strength and linear stability of the chain. $\pi$-bonds having nodes at the atomic sites behave as if they are 1D-nearly free electron system with an effective mass, $m^{*} \sim m_e$ (free electron mass) and mediate long ranged Friedel oscillations.\cite{yumur} Unusual geometric forms and emerging properties of CACs have been revealed\cite{sefaet} and freestanding CACs were produced\cite{iijima} from graphene flakes using high energy transmission microscope (TEM). Theoretically, it is also shown that CACs can be produced by stretching a graphene nanoribbon in the plastic deformation range.\cite{topsakal} Much recently, polyene consisting of 44 carbon atoms have been produced.\cite{naturechem} In an earlier experimental study, carbon adatoms and segments of carbon atomic chains were observed using TEM and attributed to vacuum contamination.\cite{nature} Since free carbon atomic chains did not form by themselves to exist as contamination, reported TEM images and video taken at finite temperature present strong evidence for our theoretical predictions.

\section{Method}

The growth mechanism we are predicting is accurately described by first-principles calculations based on Density Functional Theory (DFT) combined with ab-initio, finite temperature molecular dynamics calculations. The state-of-the art spin-polarized, first-principles plane wave calculations within DFT\cite{vasp} are carried out using projector augmented-wave (PAW) potentials\cite{paw} and local density approximation (LDA).\cite{lda} PAW potential with small core radius of 1.1 \AA~is close to all electron treatment and hence better represents C-C bond, as well as magnetic interactions in graphene+C$^{*}$ systems.\cite{kresse} In addition, high cutoff assures convergence of energies even if the sizes of superlattice are varied for different systems. We also performed calculations with generalized gradient approximation (GGA)\cite{gga} with and without van der Waals (vdW) correction\cite{grimme} for the sake of comparison with previous studies. All structures are treated within the supercell geometry, where the distance larger than 11 \AA~between any two C atoms in different cells is assured. A plane-wave basis set with kinetic energy cutoff of 900 eV is used to achieve a high precision.\cite{vasp} Brillouin zone (BZ) is sampled in the \textbf{k}-space within Monkhorst-Pack scheme,\cite{monk} where the convergence of total energy and magnetic moments with respect to the number of \textbf{k}-points in BZ are carefully tested. All atomic positions and lattice constants are optimized by using the conjugate gradient method, where the total energy and atomic forces are minimized. The convergence for energy is chosen as 10$^{-5}$ eV between two consecutive steps, the maximum Hellmann-Feynman forces acting on each atom is less than 0.04 eV/\AA{} upon ionic relaxation and the pressure is less than 1 kBar. The dipole corrections\cite{dipole} to the total energy of CAC($n$)+graphene complex is, on the average, +49 meV. In ab-initio MD calculations, the time steps are taken 2 fs and the systems are normalized at every 40 time steps.

The binding energies of adsorbates (carbon atom or CACs or other foreign atoms and molecules, such as H, Li, Co, Ti, Au and H$_2$) are calculated using the expression, $E_{b}$ = $E_{T}[Graphene]$ + $E_{T}[adsorbate]$ - $E_{T}[adsorbate+graphene]$ in terms of the ground state total energies of bare graphene $E_{T}[Graphene]$, adsorbate $E_{T}[adsorbate]$ and adsorbate+graphene complex $E_{T}[adsorbate+graphene]$. These total energies are calculated in the same unit cell using the optimized structures.

\section{Growth of carbon atomic chains on graphene}

\subsection{Carbon adatoms on graphene}

The adsorption of single carbon adatom (denoted as C$^{*}$ in the rest of the paper), which is the precursor of the growth of CACs, is treated within the periodic boundary conditions: One C$^{*}$ is assumed to be adsorbed to each $(n \times n)$ supercell of graphene resulting in a uniform coverage of one adatom per $2n^{2}$ carbon atoms in the supercell, namely $\Theta=1/2n^{2}$. Carbon adatoms favor to be bound to the bridge site, that is above the center of any C-C bond of the graphene honeycomb structure.\cite{prl} We calculated a rather strong binding energy\cite{jap} of $\sim$ 2.3 eV. We found that graphene uniformly covered by C$^{*}$ is stable for $\Theta > 1/8$. Adsorption of carbon atom on graphene\cite{hashi} and graphite surface\cite{carbon,Teobaldi} was also investigated from the first-principles. Recently, an analysis of binding energy, electronic and magnetic structures as a function of the coverage, $\Theta$ showed that carbon adatoms give rise to interesting and long ranged electronic and magnetic properties.\cite{jap} Not only C, but also other Group 4A elements, Si and Ge adatoms adsorbed to graphene\cite{canAPL} display behaviors similar to those of C$^*$. Additionally, effects of C$^{*}$ on other recently synthesized monolayer honeycomb structures are also examined.\cite{canBN,canMoS2}

\begin{center}
\begin{figure*}
\includegraphics[width=17cm]{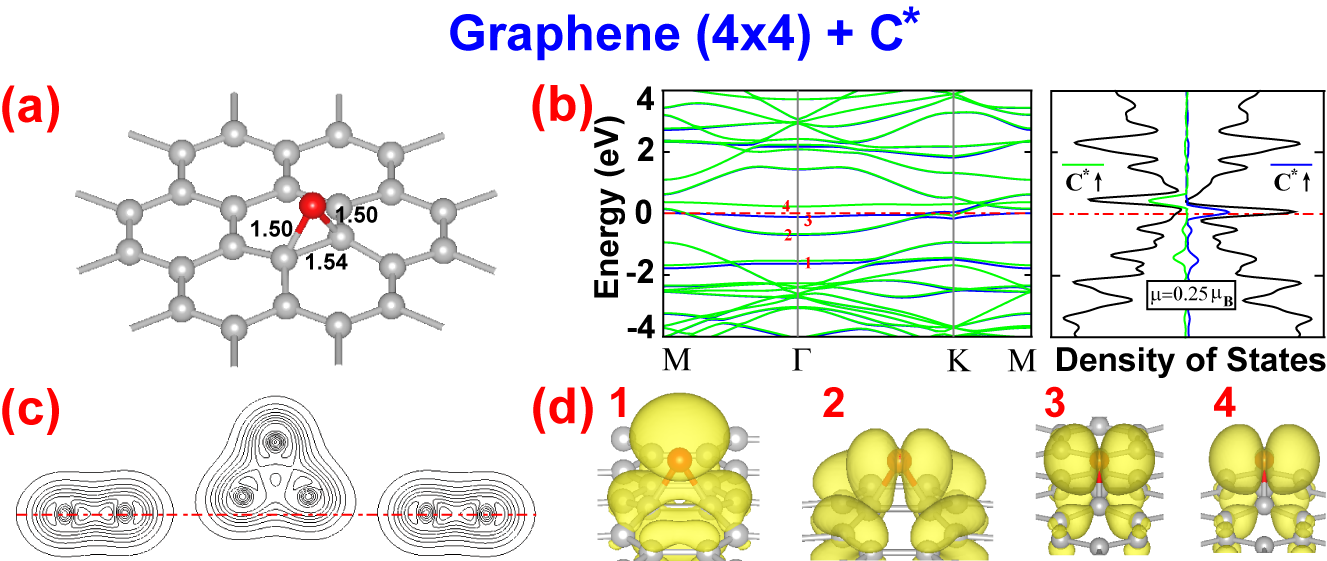}
\caption{(Color online) Adsorption geometry, electronic energy structures, and total and state charge densities of graphene+C$^*$. (a) Atomic structure: Single carbon adatom, C$^*$, (shown by the red ball) is adsorbed periodically to every $(4\times4)$ supercell of graphene (shown by grey balls) corresponding to a uniform coverage of $\Theta=1/32$. (b) Energy band structure of graphene+C$^{*}$ together with spin dependent total density of states (TDOS) and states projected to C$^*$. The zero of energy is set to the Fermi energy, $E_F$, shown by red dash-dotted lines. Spin-down and spin-up bands near $E_F$ are shown by green and blue lines, respectively. C$^*$ driven specific bands are indicated by $\sharp$1-$\sharp$4. (c) Counterplots of total charge density with contour spacings of 0.035 electrons/\AA$^{3}$. Since the density of C-C bond underlying C$^*$ is decreased, the bond is weakened and becomes longer than other C-C bonds of graphene. (d) Isosurfaces of specific states driven from C$^{*}$ as indicated by numerals, $\sharp$1-$\sharp$4 in the band structure. Isosurface values in all state charge densities are taken as $2\times10^{-5}$ electrons/\AA$^{3}$. Bands near -2 eV is formed from dangling $sp^2$-orbital of C$^*$ in $\sharp$1. Orbitals forming bridge bonds with underlying carbon atoms of graphene are clearly seen in $\sharp$2. Spin-up band originating from $\pi$-orbital of C$^*$ indicated by $\sharp$3 is shown by blue line.  This band crosses $E_F$ and attributes metallization and magnetization to the graphene+C$^*$ complex. Flat spin-down band indicated by $\sharp$4 have similar orbital character and is located just above $E_F$. Because of spin-polarization near $E_F$ graphene+C$^*$ has 0.25 $\mu_B$ (Bohr magneton) per ($4\times4$) supercell.}
\label{fig:2}
\end{figure*}
\end{center}

The effects of the adsorbed carbon atoms C$^{*}$ on the electronic structure of bare graphene are revealed by the calculations of energy band structure, total (TDOS) and projected density of states (PDOS). In Fig.~\ref{fig:2} we present the band structure corresponding to a single carbon adatom C$^{*}$ adsorbed on each $(4\times4)$ supercell of graphene ($\Theta=1/32$ uniform coverage). This structure has spin-polarized bands near $E_F$ resulting in a net magnetic moment of $\mu=0.25 ~\mu_B$ (Bohr magneton) per $(4\times4)$ supercell. The bands related with the adatom are indicated at the $\Gamma$-point by numerals from ($\sharp$1 to $\sharp$4). The flat bands near -2 eV ($\sharp$1) are derived from the dangling $sp^2$-orbital of C$^{*}$. The band ($\sharp$2) is formed from the hybridization of $\pi$-orbitals of two carbon atoms of graphene and $sp^2$-orbital of C$^*$ to form C-C$^{*}$-C bridge bonds. Other bands ($\sharp$3 and $\sharp$4) near $E_F$ are derived from spin-split dangling $p$-orbital of C$^*$ perpendicular to the plane of bridge bonded C-C$^{*}$-C. Graphene+C$^*$ complex achieves magnetic moment due to this band. When contrasted with the electronic structure of CAC later in the text one comprehends differences between C$^*$ and CAC.

\begin{center}
\begin{figure}
\includegraphics[width=8.25cm]{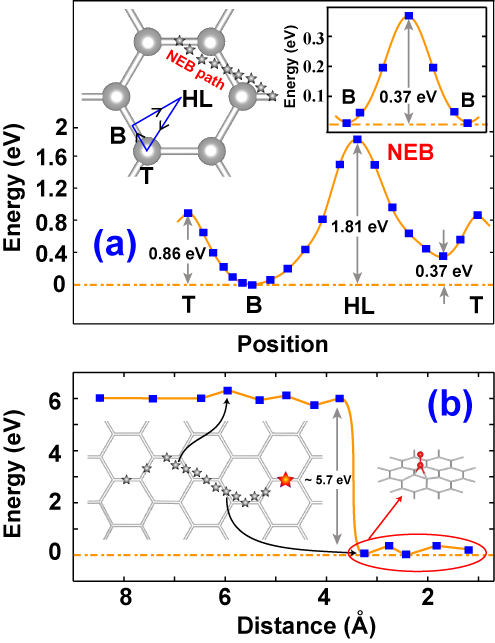}
\caption{ (Color online) (a) Energy variation of a single isolated carbon adatom,  C$^{*}$, moving along the special directions of graphene honeycomb structure. Each square corresponds to the minimum total energy of C$^{*}$ at a fixed $x$- and $y$-lateral position, but its height $z$ together with all atomic positions of graphene are optimized. \textbf{B}, \textbf{T} and \textbf{HL} indicate bridge, top and hollow sites, respectively. The migration path of a single C$^{*}$ calculated by NEB is shown by small stars on a hexagon. The energy variation on this NEB path between two adjacent \textbf{B}-site is shown by inset. (b) The interaction energy between two carbon adatoms on graphene; one is initially adsorbed at a bridge site (shown by big gold/dark star), the other moves on the path of minimum energy barrier. Within the adatom-adatom distance of 3.25 \AA~ C$_2$ is formed at the positions indicated by small stars.}

\label{fig:3}
\end{figure}
\end{center}

\subsection{Migration of carbon adatoms and chain formation, T=0}

\begin{center}
\begin{figure*}
\includegraphics[width=14cm]{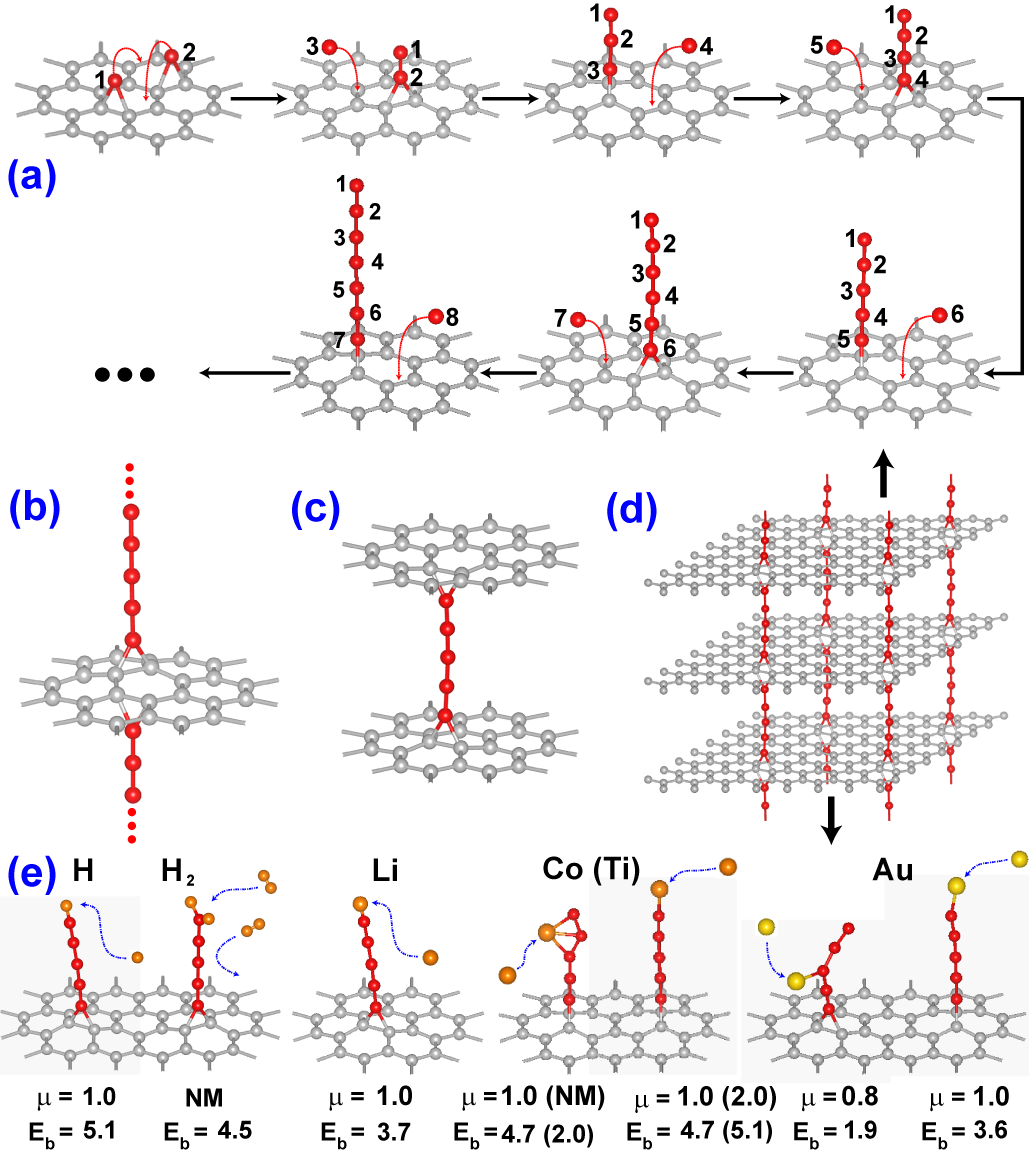}
\caption{(Color online) (a) Sequential growth of a CAC consisting of seven carbon atoms starting from two carbon adatoms at close proximity. Once the adatom-chain distance becomes smaller than a threshold distance, the adatom is inserted to the chain from the bottom end. (b) CACs can grow favorably at both sides of graphene flake. (c) Both ends of a CAC can favorably be capped by graphene flakes. (d) A stable and novel nanostructure consisting of several single layer-graphene flakes and CACs between them as pillars. (e) Sidewise approach, but head-on adsorption of H, H$_2$, Li, and sidewise as well as head-on adsorption of Co (Ti) and Au atoms (orange/medium balls) to CAC with calculated binding energy $E_b$ in eV, magnetic moment $\mu$ in $\mu_B$. NM stands for nonmagnetic state.}
\label{fig:4}
\end{figure*}
\end{center}

How carbon adatoms can migrate and form clusters on graphene can be explored through two complementary analysis. First, the energy barrier to the diffusion of a single C$^{*}$ is calculated by the NEB\cite{NEB} (nudged elastic band) method to be 0.37 eV as shown in Fig.~\ref{fig:3}(a). This barrier is, however, modified at the proximity of a second C$^{*}$. Therefore, in addition to the above analysis, the interaction between two carbon adatoms is investigated as one C$^{*}$ approaches another C$^{*}$ on a minimum-energy path as shown in Fig.~\ref{fig:3}(b). At 0 K, while this energy barrier hinders C$^{*}$ from diffusion, it is lowered as two C$^{*}$s become closer and eventually collapses totally with the onset of strong C$^{*}$-C$^{*}$ coupling. Hence, when the distance $l_{C^{*}-C^{*}}$ becomes within a threshold distance of 3.25 \AA, two carbon adatoms form CAC(2), which is attached perpendicular to the plane of graphene at the bridge site. This way, the nucleation of a CAC starts as shown in Fig.~\ref{fig:4}(a).

Chain formation of carbon adatoms at 0 K can continue once a third carbon adatom is placed at a close proximity of CAC(2) within a threshold distance. This time, through a concerted process, CAC(2) leaves its position and is attached on top of a C$^{*}$ at close proximity to form a CAC consisting of three carbon adatoms. Even more remarkable is that the chain continues to grow when the same process is repeated; each time one carbon adatom is inserted to a CAC from the bottom and hence the segment grows by one carbon atom at a time. Since the cohesive energy of a carbon atom in the infinite CAC is $\sim$ 7.8 eV, a graphene+CAC($n+1$) complex gains energy by $\Delta E$ when a CAC($n$) is united with the single C$^{*}$. We found that $\Delta E$ depends on whether $n$ is even or odd, as well as on the value of $n$, and $\Delta E \rightarrow \sim$ 5 eV as CAC becomes very long. This substantial energy gain becomes the driving force of the growth process. Sequential growths of CACs are revealed from our calculations and summarized schematically in Fig.~\ref{fig:4}(a) until a string of seven-atoms grows perpendicularly on graphene. We did not pursue further, since calculations quickly become extensive.

Free standing CACs exhibit interesting even-odd disparity depending on the number of carbon atoms, $n$ is even or odd.\cite{senger} While freestanding CACs with odd $n$ are spin-unpolarized but those with even $n$ have magnetic moment of $\mu$=2 $\mu_B$, they become spin-unpolarized when attached to graphene, no matter what $n$ is. Freestanding  CACs are linear and have cumulene type structure with nearly uniform double bonds when they have free ends or their end atoms are passivated by two hydrogens, whereas passivation of end atoms by single or triple hydrogen atoms lead to polyene structure with alternating long "single" and short "triple" bonds. When attached to graphene at the bridge site with two bonds, CACs favor cumulene like structure with slightly alternating bonds and with a different kind of even-odd disparity. For even $n$, the C-C bond of graphene underlying CAC is relatively shorter than that of CAC($n$) with odd $n$ as we will discuss in Sec.III-D. This situation is reflected to the binding energy and electronic energy structure near the band gap. The binding energies of free standing CAC($n$)s to graphene calculated in a (6x6) graphene supercell including the dipole correction are, respectively, 2.32, 2.81, 0.81, 2.42, 0.74, 2.11, 0.87 eV for $n=$ 1, 2, 3, 4, 5, 6, 7. As seen, for $n > 2$ the binding energy of a CAC with even $n$ is stronger than that with odd $n$.

A CAC grown perpendicular to graphene is strained if it is bowed; the resulting strain increases its energy with increasing curvature or bent angle. Moreover, stable CACs can also grow at both sides of the graphene plane as in Fig.~\ref{fig:4}(b). For example, the process of attaching a CAC(3) to the other side of a graphene+CAC(5) complex is exothermic by 0.54 eV. This energy would raise to $\sim$ 2 eV if a CAC with even $n$ were attached to the second surface.  It is energetically exothermic if the free end of a CAC is capped by another graphene flake as in Fig.~\ref{fig:4}(c). For example, the process of capping a graphene+CAC(5) complex by another graphene sheet is exothermic by 0.61 eV. From the combination of Fig.~\ref{fig:4}(b) and Fig.~\ref{fig:4}(c), one can derive novel structures consisting of several single layer graphene flakes having CACs between them as pillars as shown in Fig.~\ref{fig:4}(d).

\begin{center}
\begin{figure}
\includegraphics[width=8.25cm]{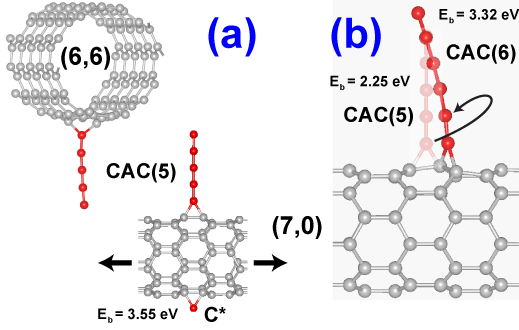}
\caption{(Color online) Growth of CACs on armchair and zigzag single wall carbon nanotubes (SWNT). (a) Optimized structures of stable CAC(5) grown on (6,6) armchair and CAC(5) and C$^*$ grown on (7,0) zigzag SWNTs. (b) Growth of CAC(6) from existing CAC(5) and C$^*$ at close proximity. Calculated binding energies $E_b$ is indicated.}
\label{fig:4p}
\end{figure}
\end{center}

Present calculations confirm the fact that surfaces of carbon nanotubes can become chemically more active as compared to flat graphene.\cite{oguz,review} For example, the binding energy of C$^*$ on (7,0) zigzag carbon nanotube is calculated to be 3.55 eV, which is approximately 1.2 eV stronger than that on bare graphene. This situation suggests that CACs can grow favorably on carbon nanotubes. In Fig.~\ref{fig:4p} (a) we show that CAC(5) can be stable on (6,6) armchair and (7,0) zigzag single wall carbon nanotubes (SWNT). The binding energy of CAC(5) is $\sim$ 2.25 eV, which is relatively stronger than that on graphene. In Fig.~\ref{fig:4p} (b) we show that the growth of CAC(6) from existing CAC(5) and C$^*$ at close proximity, which reminiscent of the growth process of CACs on graphene. In the present case, owing to the curvature of SWNT the threshold distance between CAC(5) and C$^*$ to form CAC(6) needs to be shorter as compared to the threshold distance on graphene. Again, the binding energy of CAC(6) on (7,0) SWNT is higher than that of CAC(6) on graphene and is found to be 3.32 eV.

Generally, graphene surface is not active chemically. Its activity can be enhanced through the adsorption of specific adatoms. Here we showed that the chemical activity of graphene can be promoted through CACs attached to it. In fact, as outlined in Fig.~\ref{fig:4}(e), the free end of a CAC is extremely attractive for foreign atoms, whereby graphene can attain interesting functionalities. Here we consider only H, Li, Co, Ti and Au as an example to demonstrate the enhanced chemical activity through CACs. For example, a hydrogen atom approaching sidewise to CAC jumps up and caps CAC's free end. Because of the single bond between H and CAC's free end, the structural morphology of CAC undergoes a change and the system attains a magnetic moment of 1 $\mu_B$ due to unpaired electron. While it cannot be bound sidewise, H$_2$ molecule approaching CAC's free end dissociates and forms two C-H bonds. Lithium is also adsorbed to the free end and attains 1 $\mu_B$ magnetic moment. Cobalt (and Titanium) atoms can be bound to CAC either sidewise or head-on with relatively strong binding energy and can make the system magnetic. Gold atom, which is known to have rather weak interaction with graphene, engages in strong binding with CACs. Especially, the head-on chemisorption may be useful for the stable connection of graphene with gold electrodes.

Questions whether CAC formation on graphene is the minimum energy structure is addressed by carrying out calculations of various carbon clusters consisting of 2, 3 and 4 atoms on graphene. We carried out calculations using LDA\cite{lda} and GGA\cite{gga} with and without vdW\cite{grimme} correction. The interaction between C adatoms and graphene involves chemical short range interaction and long range vdW interaction. While GGA fails to predicts the interlayer interaction, the interlayer distance (of graphite and MoS$_2$) as well as the binding energies of several adatoms, LDA results are in fair agreement with experimental values.\cite{lda-gga} Moreover, LDA is known to include part of vdW interaction.\cite{VdW}

\begin{center}
\begin{figure}
\includegraphics[width=8.25cm]{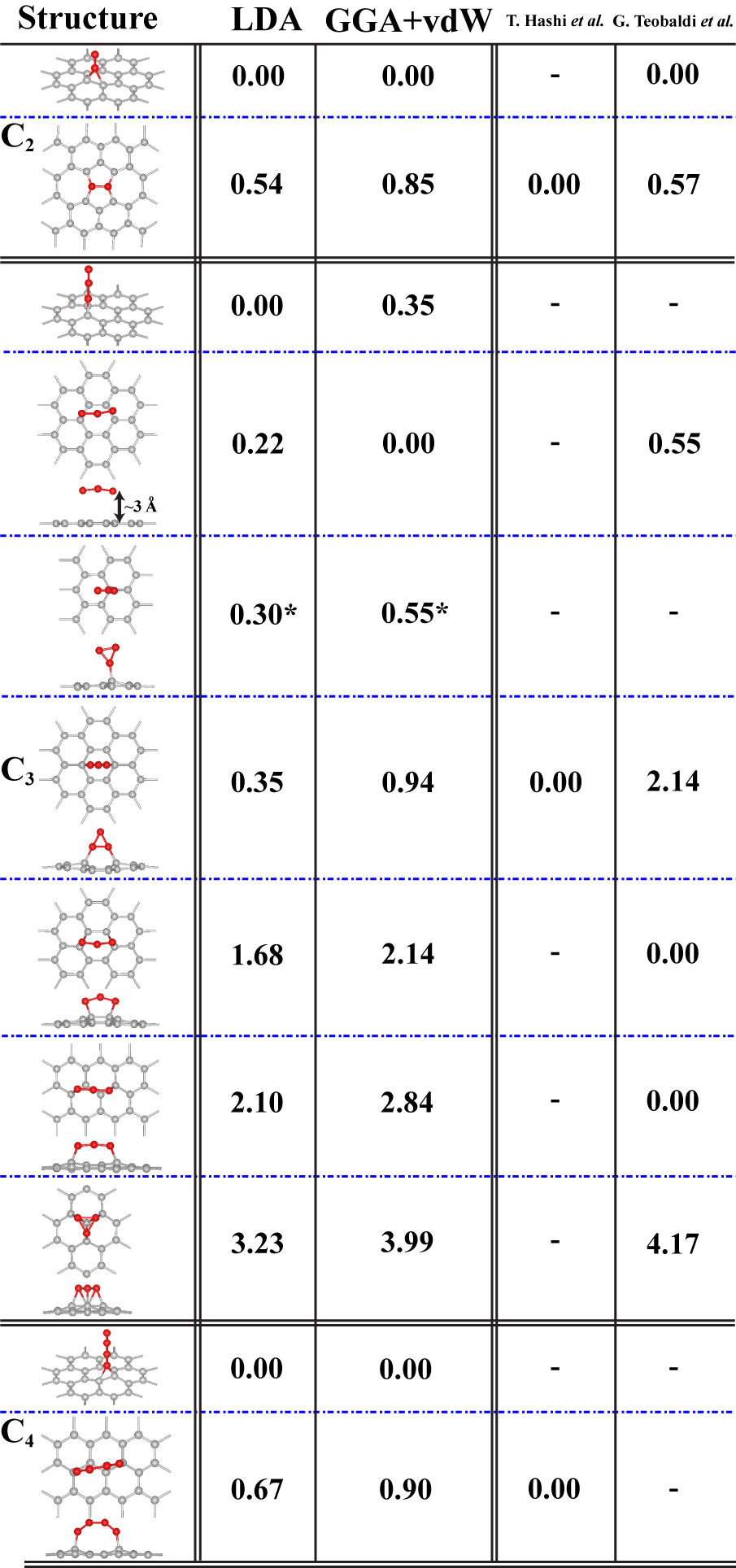}
\caption{(Color online) Comparison of energetics of CAC($n$) and various configurations of carbon atoms C$_n$ for $n$=2, 3, 4 (first column) calculated using LDA (second column), GGA+vdW (third column) methods in this work and in other works by T. Hashi \emph{et al.}\cite{hashi} (fourth column) and Teobaldi \emph{et al.}\cite{Teobaldi} (fifth column). For each $n$, the zero of energy is set to the structure having lowest energy and the energy differences of other structures with respect to the lowest energy structure are indicated in units of eV. Graphene surface is described by honeycomb structure made by grey/light balls and carbon adatoms are red/dark balls. All structures and energetics presented in this figure is nonmagnetic except a single geometry denoted by $^{*}$.}
\label{fig:5}
\end{figure}
\end{center}

Our results are presented in Fig.~\ref{fig:5} and also compared with relevant studies by Hashi \emph{et al.}\cite{hashi} and Teobaldi \emph{et al.}\cite{Teobaldi}. The contribution of vdW interaction to the binding is not large as compared to that of chemical interaction which is underestimated by GGA. For example, the vdW interaction between C* adatom and graphene is only 180 meV. Therefore, the chemical interaction dominates the binding in the present case. While the GGA binding energy of C$^*$ is calculated to be 1.4 eV, the LDA binding energy is 2.3 eV. The binding energy of C$^*$ on the surface of graphite is calculated using GGA+vDW to be 1.35 eV.\cite{Teobaldi} While GGA alone does not yield the formation of the odd numbered CACs, GGA+vdW calculations predict the formation of all chains with relatively smaller binding energy and hence confirm the LDA results.

Hashi \emph{et al.}\cite{hashi} carried out LDA calculations on carbon adatoms adsorbed to graphene using ultrasoft pseudopotential.\cite{usp} Their supercell is rectangular and consists of $128$ carbon atoms. Brillouin Zone is sampled by four \textbf{k}-points and kinetic energy cutoff is taken as $340$ eV. Spin polarization is not included in their calculations. Teobaldi \emph{et al.}\cite{Teobaldi} studied the carbon adatom adsorption on the surface of graphite slab. They used GGA+vdW and ultrasoft potential with a core radius of 1.8 \AA ~and cutoff potential of $286.7$ eV. They worked on ($4\times4\times3$) graphite slab, but sampled BZ by $4\times4\times1$ \textbf{k}-points. Here, to be consisted with other works, we carried out various cluster calculations on the ($4\times4$) supercell of graphene with \textbf{k}-point sampling of 7x7x1. In case of $n$=2, both LDA and GGA+vdW calculations find that the lowest energy structure is CAC($2$). Teobaldi \emph{et al.}\cite{Teobaldi} predict that CAC($2$) on graphite surface is also the lowest energy structure. Our results as well as that of Teobaldi \emph{et al.}\cite{Teobaldi} disagree with that of Hashi \emph{et al.}\cite{hashi} finding a different geometric structure energetically most favorable. In the case of C$_3$, LDA predicts that CAC(3) perpendicularly attached to graphene is energetically the most favorable, whereas GGA+vdW predicts that CAC(3) $\sim 3 $~\AA~ above the surface of graphene has the lowest energy.

\subsection{Interaction between CAC's}

\begin{center}
\begin{figure*}
\includegraphics[width=14cm]{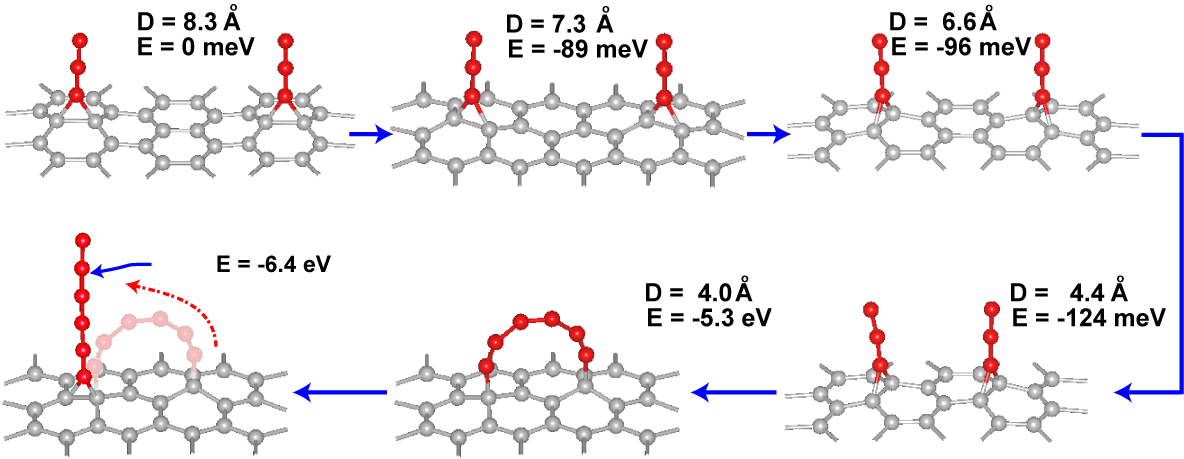}
\caption{(Color online) Variation of the total energy between two CAC($3$)s grown on graphene as a function of the spacing $D$. The total energy corresponding to $D=8.3$ \AA~is taken to be zero. The energy of graphene+CAC($3$)+CAC($3$) lowers (i.e the system gains energy) as $D$ decreases. Eventually, two CACs unite to form a handle when $D$ is smaller than a threshold distance. The total energy is further lowered when the handle is transformed to a linear chain.}
\label{fig:6}
\end{figure*}
\end{center}

Not only CAC($n$) and C$^*$ unite to form CAC($n+1$), but also two segments at close proximity, CAC($n$) and CAC($n^{'}$) can unite to make a handle like structure. Eventually, handle can transform to linear CAC($n+n'$) resulting in an energy gain. Here we examine the interaction between two CACs at close proximity at T=0 K. We first consider the interaction between two short CACs, namely CAC($n$) and CAC($n^{'}$) with $n=n^{'}=3$ as a function of spacing $D$ between them. In Fig.~\ref{fig:6}, we show the variation of the total energy for different $D$s between these CACs. For each $D$, graphene and attached CACs are relaxed. By taking the total energy of two CACs at the spacing, $D$=8.3 \AA~ as the zero of energy, the energy gets lowered when $D$ decreases. Eventually two CACs unite to form a handle (both ends attached to graphene) with an energy $E \sim$ -5.3 eV. The energy is further lowered (namely the system becomes more energetic) by $\sim$ -1.1 eV, if a handle is transformed into a linear CAC($n=6$). Similar calculations are also performed for $n$=4, $n^{'}$=2 [for CAC(6)] and $n$=4, $n^{'}$=3 [for CAC(7)]; both confirm that a single long CAC is energetically $\sim 5$ eV more favorable than two noninteracting short CACs. Using ab-initio temperature dependent molecular dynamics calculations, we also show that the unification process of two CACs at close proximity to form a longer CAC is speeded up at elevated temperatures.

\subsection{Electronic structure of CACs grown on graphene}

\begin{center}
\begin{figure}
\includegraphics[width=8.25cm]{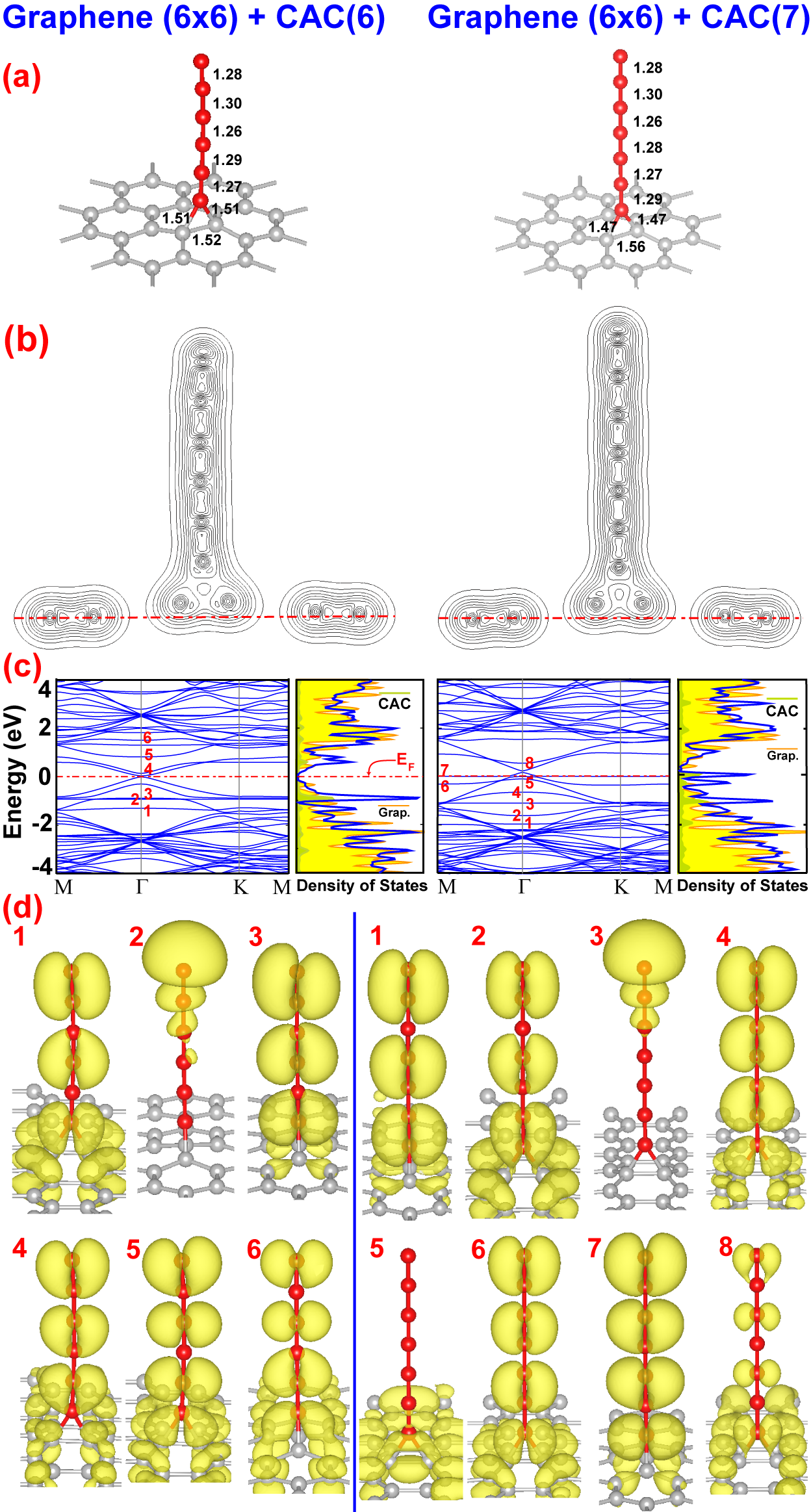}
\caption{ (Color online) Electronic energy band structure and charge densities of of graphene+CAC(6) and graphene+CAC(7) complexes. (a) Atomic structures. Lengths of various C-C bonds (in \AA) are indicated. Contour plots of total charge density in a perpendicular plane passing through C-CAC-C plane of bridge bond are also shown. Contour spacings are 0.035 electrons/\AA$^3$. (c) Electronic energy bands and total density of states (shown by blue lines) of graphene+CAC(6) and graphene+CAC(7) folded to the Brillouin zone of $(6\times 6)$ supercell. Specific bands are labeled by numerals from $\sharp$1 to $\sharp$8. States projected to CAC (shown by green lines) and total density of states of bare graphene (shown by orange lines) are also indicated for the sake of comparison. (d) Isosurfaces of charge densities of selected states indicated by numerals from $\sharp$1 to $\sharp$8 in (c). Isosurface values are taken to be $2\times10^{-5}$ electrons/\AA$^{3}$. States having charge density localized at CAC give rise to flat bands. Dispersive bands originates from states, which mix with the graphene states.}
\label{fig:8}
\end{figure}
\end{center}

The electronic band structure, total and CAC projected densities of states, total charge densities and isosurfaces of charge densities of specific states of graphene+CAC($n$) complexes are calculated for optimized structures. In Fig.~\ref{fig:8} we present our results for graphene+CAC(6) and graphene+CAC(7) complexes, which are calculated using (6x6) supercell of graphene. The contour plots of the total charge densities clarify differences in the bonding configurations of CACs with odd and even $n$. Two bonds of CAC(6) with graphene and C-C bond of graphene below CAC have almost equal lengths, namely 1.51 and 1.52 \AA. The situation is, however, different for CAC(7), which has relatively weaker binding energy with graphene. The length of the C-C bond of graphene below CAC is 1.56 \AA~ and hence it is relatively longer than two CAC-C bonds of 1.47 \AA. The same trend is found also for graphene+CAC($n$) complexes for $n$=4 and 5.

The analysis of the electronic structure of graphene+CAC(6) and graphene+CAC(7) are presented in Fig.~\ref{fig:8}. Owing to the band folding of (6x6) supercell the valence and conduction bands of graphene+CAC(6) and CAC(7), which are derived from $\pi$- and $\pi^*$-orbitals graphene cross at the $\Gamma$-point. Flat bands of CAC(6) ($\sharp$1, 2, 3) derived from CAC with minute mixing with graphene orbitals occur below $E_F$ at $\sim$-1 eV and give rise to a sharp peak in TDOS. Bands $\sharp$4, 5 and 6 above the Fermi level have increased mixing with graphene orbitals and hence increased dispersion. The bands of CAC(7) ($\sharp$1, 2, 3, 4) occur below the Fermi level and give rise to two peaks in TDOS below $E_F$. Band $\sharp$7 pins the Fermi level below the energy, where graphene $\pi$- and $\pi^{*}$-bands cross and leads the metallization of the system. The peak at $E_F$ is due to the band $\sharp$7. Similar situation occur for CAC(4) and CAC(5). This is one of the well known even-odd disparity characteristics of CAC, namely for even $n$ the states localized at CAC occur $\sim$-1 eV below the Fermi level, whereas for CAC with odd $n$ similar localized states also appear near the Fermi level. We finally note that while free CACs with odd $n$ are spin-unpolarized, but those with even $n$ have magnetic moment of $\mu$=2 $\mu_B$, they become spin-unpolarized when attached to graphene, no matter what $n$ is.

\subsection{Irregular growth}

\begin{center}
\begin{figure}
\includegraphics[width=8.25cm]{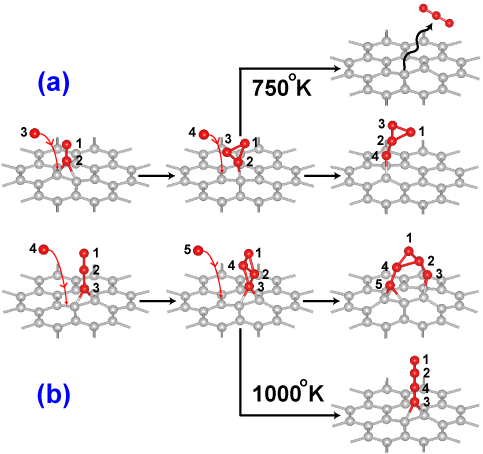}
\caption{ (Color online) Schematic description of irregular growth of CACs, when the distance between a CAC and C$^*$ becomes accidentally close to $\sim$ 1.50 \AA. (a) C$^{*}$ and C$_2$ form a triangular cluster, which is attached to the top site from its corner. Thereafter CACs can grow regularly if additional C$^*$ approaches the triangular cluster not any closer than 2.1 \AA. At 750 K, the triangular cluster is removed from graphene and subsequently it transforms to freestanding linear CAC in vacuum. (b) Irregular growth of a quadrangle formed from CAC(3) and C$^*$ and adsorbed at the bridge site under similar conditions as (a). Irregularity may continue in the next growth step, but at 1000 K quadrangle cluster can transform to the linear CAC(4).}

\label{fig:9}
\end{figure}
\end{center}

Finally, apart from the above regular sequences, irregular growth may take place, when a C$^{*}$ accidentally gets as close as $\sim$1.50 \AA~to an existing CAC(2) (or CAC(3)). At the end a tilted triangle (quadrangle) of carbon atoms can form, which, in turn, is bound to the top (bridge) site from one corner and have nonmagnetic state. These are irregular and nonequilibrium processes, since they may occur even if these structures are not energetically favorable. For example, graphene with triangular (quadrangular) cluster is 0.36 eV (0.92 eV) less energetic than the linear CAC(3) (CAC(4)). At high temperature, while a quadrangular cluster changes to a linear CAC attached to graphene, triangular one is first detached, later changes to three-atom chain in the vacuum. Irregular forms of CAC growth are shown in Fig.\ref{fig:9}.

\subsection{Analysis at finite temperature}

The coexistence of two C$^{*}$ adatoms or one C$^{*}$ and a CAC within the threshold distance was a prerequisite for the growth process at T=0 K. Despite an energy barrier of $Q=$0.37 eV a single C$^{*}$ can migrate readily above room temperature with a diffusion constant, $D= \nu a e^{-Q/k_{B}T}$, to be within the threshold distance of another C$^{*}$ (or another CAC). Thereafter, CAC(2) (or a longer CAC) can grow. Here, $a$ is the lattice constant. The characteristic jump frequency is estimated to be $\nu =7.5 \times 10^{12}$ $s^{-1}$ from the phonon calculations\cite{jap} of graphene+C$^*$. Ab-initio molecular dynamics (MD) calculations carried out with fixed number of atoms at finite temperature corroborate the above mechanism of growth revealed by structure optimization at T=0 K. High temperature behavior and the stability of graphene+CAC($n$) with $n$=2-7 are also investigated by holding them at various temperatures ranging from 400 K to 1600 K for a number of time steps ranging from 200 to 1000. Even if enough statistics cannot be accumulated in a several thousands steps at elevated temperatures, it becomes clear that CAC($n$) on graphene are stable and at least they cannot desorb readily at room temperature. CACs rather start to swing; occasionally either they change their adsorption sites or their free ends also attach to graphene to form handle like structures.

Here we present a few example for our analysis at finite temperature. The binding energy of CAC(4) to graphene is calculated to be 2.42 eV. It is large enough to assure the stability and to hinder desorption just above the room temperature. Ab-initio molecular dynamics calculations are carried out for graphene+CAC(4) at T=300 K, 600 K, 1000 K, 1500 K and 1600 K each for 200 step. CAC(4) swings at T=300 K, 600 K and 1000 K, but they do not form handle like structures with two ends bound to graphene. However, when the temperature is raised to T=1500 K and further, they first swing and then form handle like structure. Eventually, they wander on graphene. For T$\leq $1600 K desorption of CAC did not take place.  Nonetheless, desorption of CAC(4) from graphene could have taken place if the number of time steps were very large. On the other hand, CAC(5) on graphene can swing and form handle like structure at T=500 K. At T=1000 K, it also swings, forms handle like structure and eventually is detached from graphene surface. For graphene+CAC(7) calculations are performed initially at T=400 K for 1000 steps; later, the temperature is raised to T=600 K and calculations continued another 1000 steps. No desorption did occurred at T=600 K within 1000 time steps. As CACs are swinging they diffuse on graphene through the path, bridge-top-bridge sites.  In view of ab-initio MD calculations we draw following conclusions. (i) The binding energy of CACs with odd $n$ tend to desorb at relatively lower temperatures. Since GGA+vdW calculations yields relatively lower binding energies, the desorption temperatures predicted therefrom are expected to be lower than those of LDA calculations. (ii) At moderate temperatures CACs can swing and wander on the surface of graphene. (iii) Since C-C bond in a CAC is stronger than the bridge bond between chain and graphene, carbon atoms do not desorb from the free end of a CAC, rather whole chain is desorbed. (iv) Finally we note that the dynamics of CACs revealed from our ab-initio MD calculations are similar to the videos taken from TEM images of diffusing carbon chains on graphene.\cite{nature}

\section{Carbon adatoms at the edge of graphene}

\begin{center}
\begin{figure}
\includegraphics[width=8.25cm]{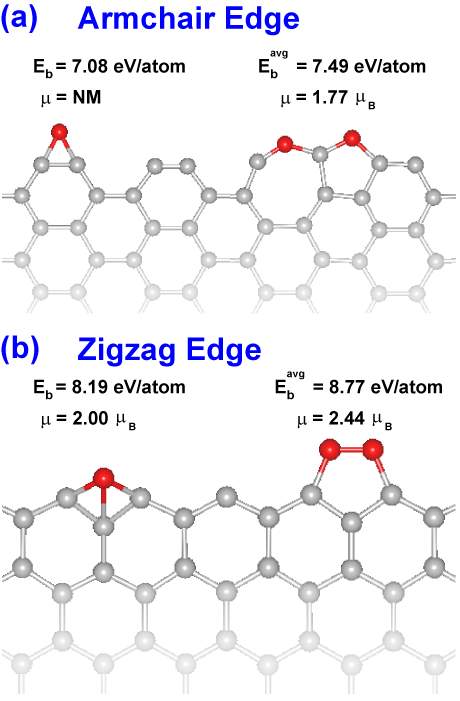}
\caption{ (Color online) Schematic description of the growth of graphene flake at the edges to expand its size.
(a) Bonding configuration of single and two carbon adatoms adsorbed to the edge of armchair nanoribbon. (b) Same for the edge of zigzag nanoribbon. Binding energies of carbon adatoms, $E_b$, and total magnetic moment, $\mu$,  are given in eV and $\mu_{B}$, respectively. NM stands for nonmagnetic state.}
\label{fig:10}
\end{figure}
\end{center}

In the above analysis favoring the growth of CACs we used periodic boundary condition, whereby
graphene sheet did not have any edge. Here the important question one has to address is whether CACs still grow on finite size graphene sheets or migrating C$^*$s prefer to fill the empty atomic sites at the edge in registry with graphene crystal. The latter case is related with the growth of graphene from edges. Earlier it was revealed that the binding energy of carbon adatom at the edge of hydrogen saturated armchair (zigzag) nanoribbons is 3.81 (4.86) eV and hence is stronger than that at the center of the ribbon (2.3-2.7 eV).\cite{jap} The bonding configuration is different from that of C$^*$ on graphene. This indicates that a single C$^*$ favors to be at the edges of graphene, unless it is already inserted to a CAC away from the edge to lower its total energy by $\sim$ 5 eV. We examined the bonding of C$^*$ at the edges of bare armchair and zigzag graphene nanoribbons. The configurations comprising single and two carbon adatoms at the edges of armchair and zigzag nanoribbons are presented in Fig.\ref{fig:10}. Here nanoribbons are used to model finite size graphene flakes. In  Fig.\ref{fig:10} (a) we show bonding configuration of single and two carbon adatoms adsorbed to the edge of armchair nanoribbon. While the cohesive energy of carbon atom in graphene is calculated as 8.98 eV, a single carbon adatom prefers to saturate two dangling $sp^2$ bonds to form a bridge bond with a binding energy of $E_b$=7.08 eV. It is much higher than the binding energy of C$^*$ on graphene. The second C$^*$ at the close proximity does not combine into CAC(2), but forms fivefold and sevenfold rings with an average cohesive energy of 7.49 eV/atom. In Fig.\ref{fig:10} (b) one faces a similar situation at the zigzag edge; a single carbon adatom saturates two $sp^2$ to form a bridge bond above the plane of nanoribbon with a binding energy of 8.19 eV. The ground state of two carbon adatoms is the pentagon formed at the edge with an average binding energy of 8.77  eV/atom. These binding energies indicate that a carbon adatom reached to the edge of a graphene flake favors to expand the size of graphene, rather than forming a CAC at the edge. However, the formation of a CAC is favored away from the edges. Whether the epitaxial growth of graphene from the edge continues from bridge bonded carbon adatoms is beyond the scope of this study.

\section{Conclusions}

In conclusion, we showed that carbon adatoms readily diffuse on graphene above room temperature and nucleate C$_2$, which subsequently grows as linear carbon chains perpendicularly attached to graphene. Similar growth processes are also shown on single wall carbon nanotubes, graphene nanoribbons, as well as on graphite surface. It is shown that through the coverage of CAC the chemical activity of graphene is enhanced and some of the physical properties are dramatically modified. The coverage of CACs and its physical and chemical properties (such as desorption and conductance) can be monitored by perpendicular and lateral bias voltage applied to graphene or by charging the system.\cite{topsakal2}

The cohesion of free carbon chains is rather strong and is comparable either with the average cohesive energy of a small cluster of diamond having $sp^3$ bonding or graphite cluster having $sp^2$ + $\pi$ bonding with large surface to volume ratio. Hence the significant energy gain provided by a carbon adatom implemented to an existing CAC or to another carbon adatom is the driving force leading to the formation of CACs. This self-assembling behavior of carbon adatoms on graphene is not only of fundamental interest, but also offers artificial nanostructures with interesting future applications. Large spacing sustained by CACs behaving like pillars between multiple graphene layers can be utilized as diverse intercalation systems. Graphene and its nanoribbons, as well as nanotubes can establish connections to other nanostructures through CACs. Specific molecules or atoms attached to CACs modify physical properties of graphene+CAC complex, which in turn can be utilized as sensors. In particular, Li atoms capping short CACs can function as a high capacity hydrogen storage medium with $\sim$10 wt\%. In summary, the growth of novel graphene+CAC complexes and their important applications promise a new perspective in graphene research.

\begin{acknowledgments}
This work is supported by TUBITAK through Grant No: 108T234. Computing resources used in this work were partly provided by the National Center for High Performance Computing of Turkey (UYBHM) under grant number 2-024-2007. Part of the computational resources has been provided by TUBITAK ULAKBIM, High Performance and Grid Computing Center (TR-Grid e-Infrastructure). SC acknowledges TUBA-Academy of Science of Turkey and DPT-State Planning Organization for partial support.

\end{acknowledgments}

\end{document}